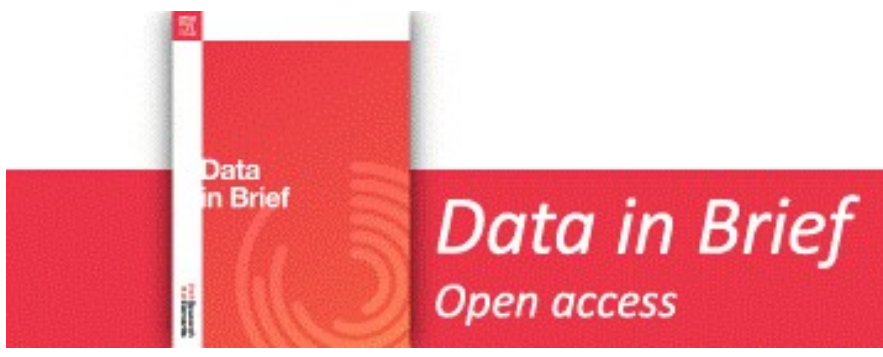

Article template

# ARTICLE INFORMATION

**Article title**

Adiabatic Theory Data on Strongly Chirped Dissipative Solitons of the Cubic-Quintic Nonlinear Ginzburg-Landau Equation

**Authors**

V. L. Kalashnikov[1,*], E. Sorokin[2,3], A. Rudenkov[1], I. T. Sorokina[1,3].

**Affiliations**

[1]Department of Physics, Norwegian University of Science and Technology, Realfagbygget, Gløshaugen, Høgskoleringen, Trondheim, NO-7491, Norway;

[2]Institut für Photonik, TU Wien, Gußhausstraße 27/387, Vienna, A-1040, Austria;

[3]ATLA lasers AS, Richard Birkelands vei 2B, Trondheim, NO-7034, Norway

**Corresponding author's email address and Twitter handle**

vladimir.kalashnikov@ntnu.no

**Keywords**

stationary-phase approximation; master diagram; dissipative soliton resonance; coherent condensates.

**Abstract**

This data article provides the datasets, symbolic derivations, and scripts used to reproduce master diagrams, stationary-phase spectra, windowed first-order coherence functions, and quantum-noise stability maps for strongly chirped dissipative solitons of the cubic-quintic complex Ginzburg-Landau equation in normal and anomalous group-delay dispersion regimes. The repository includes node-regularized normal-dispersion spectra and energies; small-parameter expansions of the branch roots; cavity-map gain-loss update relations; Airy uniformization at the normal-dispersion spectral edge; anomalous-dispersion spectra and coherence calculations; and processed tables for plotting and stability analysis. OriginLab projects are accompanied by open-format .csv/.txt numerical tables to support reuse without proprietary plotting software. Data and code repository: https://doi.org/10.5281/zenodo.22690899.

# SPECIFICATIONS TABLE

| **Subject** | Computer Sciences |
|---|---|
| **Specific subject area** | Dissipative solitons and dissipative soliton resonance in the cubic-quintic complex Ginzburg-Landau equation; Thermodynamics of Non-Equilibrium Dissipative |
| | Systems. |
| **Type of data** | MATLAB scripts (.m); Maple worksheets (.mw); OriginLab projects (.opju); open-format numerical tables (.csv and .txt); processed numerical tables; and derivation comments (.pdf). |
| **Data collection** | Original computational data generated by symbolic algebra, analytical evaluation, parameter-grid calculations, Fourier/cosine transforms, and stochastic split-step propagation. The principal scanned variables are the normalized CGLE control parameters, dispersion sign, spectral window, energy targets, and stochastic-shot parameters used for the AGD finite-time survival analysis. |
| **Data source location** | Department of Physics, Norwegian University of Science and Technology, Realfagbygget, Gløshaugen, Høgskoleringen, Trondheim, NO-7491, Norway. |
| **Data accessibility** | Repository name: ZENODO<br>Data identification number / DOI: 10.5281/zenodo.22690899<br>Direct URL to data: https://doi.org/10.5281/zenodo.22690899 |
| **Related research article** | V. L. Kalashnikov, E. Sorokin, A. Rudenkov, and I. T. Sorokina, “Strongly chirped dissipative solitons in normal and anomalous dispersion regimes”, Chaos, Solitons & Fractals **210**, 118609 (2026). |

# VALUE OF THE DATA

• The deposited symbolic derivations, Maple and MATLAB scripts, and tabulated numerical outputs allow independent reproduction of the master diagrams, NGD/AGD stationary-phase spectra, first-order coherence functions, and AGD finite-time survival maps associated with the related research article [1].

• The data separate analytical formulas from executable evaluations, enabling users to test limiting cases, vary the normalized control parameters, compare independent solvers, and benchmark numerical implementations of cubic-quintic CGLE models.

• The workflows are reusable beyond the specific examples considered here. They can support studies of strongly chirped pulses in fiber and solid-state mode-locked lasers and, after appropriate reinterpretation of coefficients and observables, related driven-dissipative nonlinear-wave and condensate models.

• OriginLab projects are accompanied by open .csv/.txt tables, so the numerical curves and maps can be reconstructed in MATLAB, Python, spreadsheet software, or other environments without an OriginLab license. The open tables retain the normalized variables and diagnostic quantities used in the manuscript.

# BACKGROUND

Strongly chirped dissipative solitons (DSs) provide an energy-scalable class of localized states in mode-locked lasers described by cubic-quintic complex Ginzburg-Landau (CGLE) models. The related research article [1] develops an adiabatic treatment for normal- and anomalous-group-delay dispersion (NGD and AGD), combining analytical existence conditions, stationary-phase spectra, master diagrams, first-order coherence functions, and a quantum-noise test of the dynamically accessible part of the AGD branch.

The calculations behind these results contain several reproducibility layers that are cumbersome to include in a physics article: symbolic reduction of the normalized equations, expansion of the cutoff branches, node and spectral-edge regularization, numerical evaluation of energy and coherence integrals, parameter-grid scans, and stochastic propagation of perturbations about an analytical AGD pulse. The present data article collects these original computational materials and exposes the intermediate formulas, scripts, and tabulated outputs needed to reconstruct the published results.

The dataset is intended for three complementary uses: direct reproduction of figures and numerical values; independent verification or benchmarking of the analytical and numerical procedures; and extension of the same normalized workflow to other dissipative-soliton parameter sets. The underlying CGLE structure also makes the data relevant to broader driven-dissipative nonlinear-wave problems, including laser systems and, where the coefficients are reinterpreted appropriately, weakly dissipative condensate models.

# DATA DESCRIPTION

The complete dataset is deposited at Zenodo (DOI: 10.5281/zenodo.22690899; https://zenodo.org/records/22690899). It contains four data classes: (i) Maple worksheets and derivation notes for the analytical reductions, cutoff-branch expansions, and AGD energy expressions; (ii) MATLAB scripts for spectra, coherence functions, cavity-map relations, Airy edge regularization, and AGD finite-time survival calculations; (iii) OriginLab project files used for publication-quality plotting; and (iv) open .csv/.txt exports of the numerical datasets so that the deposited results can be reused independently of OriginLab. Processed files retain the parameter values used to generate each curve, map, or histogram.

For direct reproducibility, Table 1 maps the principal repository files to the corresponding local equations and figures in this data article and to the associated results in the related research article [1].

*Table 1. Lookup between repository files and the corresponding equations and figures.*

| Repository file(s) | Open-format companion | Data article location | Related research article |
|---|---|---|---|
| NGD_node_regularization.mw | NGD_node_regularization.csv | Sec. 4.2, Eqs. (3)-(6) | Fig. 3; NGD node-regularized spectrum and energy |
| cutoff_branch_series.mw | cutoff_branch_series.csv | Sec. 4.3, Eqs. (7)-(13) | Fig. 2; cutoff-branch expansion about $\tilde{\chi}$ = 0 |
| cavity_map_sigma.m | cavity_map_sigma.txt | Sec. 4.4, Eqs. (14)-(16) | Master-diagram/isogain interpretation |
| NGD_Airy_uniformization.m | NGD_Airy_uniformization.csv | Sec. 4.5, Eqs. (17)-(20) | Fig. 3(a); NGD spectral-edge uniformization |
| AGD_multihorn_spectra.m | AGD_multihorn_spectra.csv | Sec. 4.6, Eq. (21) | AGD spectral profiles and coherent multi-horn structure |
| AGD_energy.mw | AGD_energy.txt | Sec. 4.7, Eqs. (22)-(24) | AGD energy scaling / master-diagram analysis |
| AGD_autocorrelation.m | AGD_autocorrelation.csv | Sec. 4.8, Eqs. (25)-(34) | Fig. 7; first-order coherence and two correlation scales |
| AGD_stability_map.m | AGD_stability_map.csv | Sec. 4.9, Fig. 1 | Fig. 8; AGD quantum-noise stability map |
| AGD_representative_profiles.m | AGD_representative_profiles.csv | Sec. 4.9, Fig. 2 | Fig. 9; representative analytical/noisy AGD profiles |

The .csv/.txt exports contain the same numerical columns used for the corresponding OriginLab plots. Parameter-grid files use the dimensionless variables defined in Eq. (2), including C, $\tilde{\chi}$, Σ, $\tilde{E}$, and δC where applicable; survival probability, perturbation norm, and spectral-correlation diagnostics are dimensionless. Temporal and spectral profile tables use the normalized coordinates $\tilde{t}$ and Ω and normalized power or spectral intensity unless a dimensional quantity is identified explicitly in the file header. The open tables can therefore be imported directly into MATLAB, Python, spreadsheet software, or other plotting environments.

The archived scripts are distributed as source rather than compiled binaries. All parameters defining a particular run are exposed in the corresponding script or worksheet, while the open numerical tables permit inspection and plotting even when Maple or OriginLab is unavailable.

No minimum MATLAB or Maple release is asserted here because the deposited material is source-level and the reusable numerical outputs are software-independent; users should consult the repository file headers/README for the exact environment used for the archived calculations.

# EXPERIMENTAL DESIGN, MATERIALS AND METHODS

All data were generated computationally. The reproducibility workflow used for the deposited files is summarized below; the following subsections provide the equations and implementation details for each stage.

Step 1 - Model definition and normalization. Define the cubic-quintic CGLE coefficients, choose the dispersion sign, and convert dimensional variables to the normalized coordinates of Eq. (2).

Step 2 - Analytical branch selection. Evaluate the algebraic branch roots and admissibility conditions on the chosen (C, $\tilde{\chi}$, Σ) parameter point or grid, rejecting points that violate the physical sign and reality constraints.

Step 3 - NGD data generation. Construct the stationary-phase NGD spectrum, apply the node regularization and, when required, the Airy edge uniformization, and integrate the spectrum to obtain the corresponding normalized energy. The small-$\tilde{\chi}$ branch coefficients are generated symbolically from the recurrence relations.

Step 4 - AGD data generation. Introduce the positive AGD scale D_a, evaluate the non-truncated stationary-phase spectrum and the closed-form/integral energy representation, and generate parameter scans approaching or detuning from the AGD-DSR locus.

Step 5 - Coherence data. Apply the finite spectral window used for observation/plotting, calculate the first-order autocorrelation by cosine transform, normalize it to obtain g^(1), and extract the short window-controlled and long core-controlled correlation scales.

Step 6 - Figure and table export. Store the parameter values and evaluated curves in the plotting datasets; export the same numerical columns in OriginLab and open .csv/.txt forms.

Step 7 - AGD finite-time survival data. Reconstruct the analytical AGD pulse, add independent band-limited Wigner-vacuum perturbations, propagate the linearized perturbation equation with the split-step procedure, and classify each realization using the perturbation, pulse-number, and spectral-correlation criteria described in Sec. 4.9.

Step 8 - Validation. Repeat selected calculations with refined numerical grids or alternative direct quadrature, check limiting cases against reduced-CGLE expressions, and verify that exported tables reproduce the plotted curves.

### 4.1. Normalization and notation

Throughout this work, we use the same $(1+1)$D cubic–quintic CGLE as in the related research article [1]:

$$\frac{\partial}{\partial z}a(z,t)=-\sigma a(z,t)+(\alpha-i\beta)\frac{\partial^2 a}{\partial t^2} \\ +(\kappa+i\gamma)P(t)a(z,t)-(\kappa\zeta+i\chi)P(t)^2 a(z,t). \quad (1)$$

Here, $z$ is a “longitudinal” coordinate (e.g., propagation distance in a fiber or evolution time for BEC), $t$ is a local time (“transverse coordinate”), $a(z,t)$ is a slowly-varying field amplitude, $P(z,t)=|a(z,t)|^2$ is a power. The dissipative parameters are: $\sigma$ (saturated loss), $\alpha$ (inverse squared bandwidth of a spectral filter), $\kappa$ (SAM coefficient), $\zeta$ (SAM saturation parameter). Non-dissipative parameters: $\beta$ (GDD, positive for NGD and negative for AGD), $\gamma$ (SPM), and $\chi$ (quintic phase nonlinearity). The dimensionality of the Eq. (1) variables and parameters is cataloged in Table 2.

We use the dimensionless variables:

$$\tilde{\Omega}^2=\frac{|\beta|\zeta}{\gamma}\Omega^2, \tilde{\Delta}^2=\frac{\beta\zeta}{\gamma}\Delta^2, \tilde{q}=\frac{\zeta}{\gamma}q, \tilde{P}=\zeta P, \\ \tilde{t}=\frac{\kappa}{\sqrt{|\beta|\gamma\zeta}}t, \tilde{\chi}=\frac{\chi}{\gamma\zeta}, C=\frac{\alpha\gamma}{\beta\kappa}, \Sigma=\frac{\zeta}{\kappa}\sigma. \quad (2)$$

Since $\Omega t=(\gamma/\kappa)\tilde{\Omega}\tilde{t}$, Fourier/cosine transforms are defined in $(t,\Omega)$ and then rewritten in $(\tilde{t},\tilde{\Omega})$ using the Jacobian $d\Omega=\sqrt{\gamma/(|\beta|\zeta)}\,d\tilde{\Omega}$ and the scaled spectral power.

For a real scaling in branches where $\beta\zeta<0$, replace the square root in $\tilde{t}$ by $\sqrt{\gamma\zeta|\beta|}$ and carry the sign in $\tilde{\Omega}^2$ and $\tilde{\Delta}^2$.

Table 2. Dimensions, normalized forms, and sign conventions of the variables and parameters in the cubic–quintic CGLE.

| Symbol | Dimension | Meaning | Normalized form | Comment / sign |
|---|---|---|---|---|
| $z$ | m | longitudinal coordinate | – | Dimensionless. |
| $t$ | s | local time / transverse coordinate | $\tilde{t}=t\kappa/\sqrt{\gamma\zeta\beta}$ | For real scale use $\sqrt{\gamma\zeta|\beta|}$. Not conjugate to $\tilde{\Omega}$ unless $\kappa=\gamma$. |
| $a(z,t), a(t)$ | $W^{1/2}$ | field envelope | $\tilde{a}=\sqrt{\zeta}\,a$ | Dimensionless only algebraically. For $\zeta<0$, use $\sqrt{|\zeta|}$ for a real amplitude scale and keep the sign in $\tilde{P}$. |
| $P(t)=|a|^2$ | W | power | $\tilde{P}=\zeta P$ | $sgn\,\tilde{P}=sgn\,\zeta$ for physical $P>0$. |
| $P_0$ | W | peak power | $\tilde{P}_0=\zeta P_0$ | Same sign rule as $\tilde{P}$. |
| $\phi(t)$ | | phase | – | Dimensionless. |
| $\Omega=d\phi/dt$ | $s^{-1}$ | local angular frequency | $\tilde{\Omega}^2=\Omega^2\zeta\beta/\gamma$ | $\tilde{\Omega}=\Omega\sqrt{\zeta|\beta|/\gamma}$ only after choosing a branch of the square root. |
| $\Delta$ | $s^{-1}$ | frequency cut-off | $\tilde{\Delta}^2=\Delta^2\zeta\beta/\gamma=-\tilde{q}$ | Since $\Delta^2=-q/\beta$, $sgn\,\tilde{\Delta}^2=sgn(\beta\zeta)$ for physical $\Delta^2>0$. |
| $q$ | $m^{-1}$ | wave number / propagation constant shift | $\tilde{q}=q\zeta/\gamma$ | From $\tilde{\Delta}^2=-\tilde{q}$. |
| $\sigma$ | $m^{-1}$ | saturated net loss/gain coefficient | $\Sigma=\zeta\sigma/\kappa$ | Sign is not fixed by dimensions. If $\sigma,\kappa>0$, $sgn\,\Sigma=sgn\,\zeta$. |
| $\alpha$ | $s^2 m^{-1}$ | spectral filtering | – | Usually $>0$. |
| $\beta$ | $s^2 m^{-1}$ | group-delay dispersion coefficient | – | $\beta>0$ NGD, $\beta<0$ AGD in the paper. |
| $\kappa$ | $m^{-1}W^{-1}$ | cubic saturable-absorption coefficient | – | Usually $>0$. |
| $\zeta$ | $W^{-1}$ | SAM saturation parameter | – | The paper's Table 1 allows $\zeta\gtrless 0$. If one imposes ordinary positive saturation, retain only $\zeta>0$ branches. |
| $\gamma$ | $m^{-1}W^{-1}$ | self-phase modulation coefficient | – | Usually $>0$. |
| $\chi$ | $m^{-1}W^{-2}$ | quintic SPM coefficient | $\tilde{\chi}=\chi/(\gamma\zeta)$ | Sign depends on $\chi/\zeta$. |
| $C$ | | dissipative-to- | $C=\alpha\gamma/(\beta\kappa)$ | If $\alpha,\gamma,\kappa>0$, $sgn\,C=sgn\,\beta$. |

| Symbol | Dimension | Meaning | Normalized form | Comment / sign |
|---|---|---|---|---|
| | | conservative cubic ratio | | |
| $A$ in Eq. (11) | $m^{-2}W^{-2}$ | discriminant under $\sqrt{A}$ | $A/\gamma^2$ | $\sqrt{A}$ has the same dimension as $\gamma$. The $A$ in Eq. (17) is already dimensionless. |
| $\Theta$ | | dimensionless frequency in Eq. (14) | $\Theta^2=\tilde{\Omega}^2$ | Use $\Theta$ rather than dimensional $\Omega$ in the dimensionless ODE. |
| $Y,\Psi_{\mp}$ | | auxiliary functions in Eqs. (14)–(16) | – | Dimensionless. |

## 4.2. Quadratic expressions for the node-window expansion

Starting from the NGD stationary-phase envelope and its node-window expansion in the related research article [1], one may expand $\Xi_\chi$ about $\tilde{\Omega}_0$ and express the result as a quadratic in $Y_\chi$.

$$\Xi_\chi(\tilde{\Omega})\approx A_0(\tilde{\chi})+A_1(\tilde{\chi})Y_\chi(\tilde{\Omega})+A_2(\tilde{\chi})Y_\chi(\tilde{\Omega})^2,$$

with the *explicit* coefficient formulas

$$A_0=\Xi_\chi(\tilde{\Omega}_0),A_1=\frac{\Xi'_\chi(\tilde{\Omega}_0)}{Y'_\chi(\tilde{\Omega}_0)},A_2=\frac{\Xi''_\chi(\tilde{\Omega}_0)-A_1Y''_\chi(\tilde{\Omega}_0)}{2\left[Y'_\chi(\tilde{\Omega}_0)\right]^2},$$

and the derivatives

$$Y'(\tilde{\Omega})=\frac{4\tilde{\chi}\tilde{\Omega}}{Y},Y''(\tilde{\Omega})=\frac{4\tilde{\chi}}{Y}-\frac{16\tilde{\chi}^2\tilde{\Omega}^2}{Y^3},$$

$$Y_\chi'(\tilde{\Omega})=-4\tilde{\Omega}Y+(\tilde{\Delta}^2-2\tilde{\Omega}^2)Y'(\tilde{\Omega})-2\tilde{\Omega},$$

$$Y_\chi''(\tilde{\Omega})=-4Y-4\tilde{\Omega}Y'+(\tilde{\Delta}^2-2\tilde{\Omega}^2)Y''(\tilde{\Omega})-4\tilde{\Omega}Y'-2.$$

Let us introduce a small $\chi$-vanishing regulator $\delta_\chi=K\tilde{\chi}^2$ (e.g. $K=\frac{4}{3}\tilde{\Delta}^2$), and define the node–regularized spectrum

$$\hat{S}_{NGD}(\tilde{\Omega};\tilde{\chi})=\frac{\gamma^2\zeta|\beta|}{2\pi\kappa}S_{NGD}(\tilde{\Omega};\tilde{\chi})=\frac{2\zeta|\beta|}{\kappa^3}\frac{\sqrt{Y_\chi(\tilde{\Omega})^2+\delta_\chi^2}}{\left|A_0+A_1Y_\chi(\tilde{\Omega})+A_2Y_\chi(\tilde{\Omega})^2\right|}. \quad (3)$$

This reduces to the $\chi\to0$ node–regularized formula and captures the $\chi$-dependent node shift through $\tilde{\Omega}_0(\chi)$.

With $E=\int_{-\tilde{\Delta}}^{\tilde{\Delta}}\hat{S}\,d\tilde{\Omega}/(2\pi)$ and evenness, we have

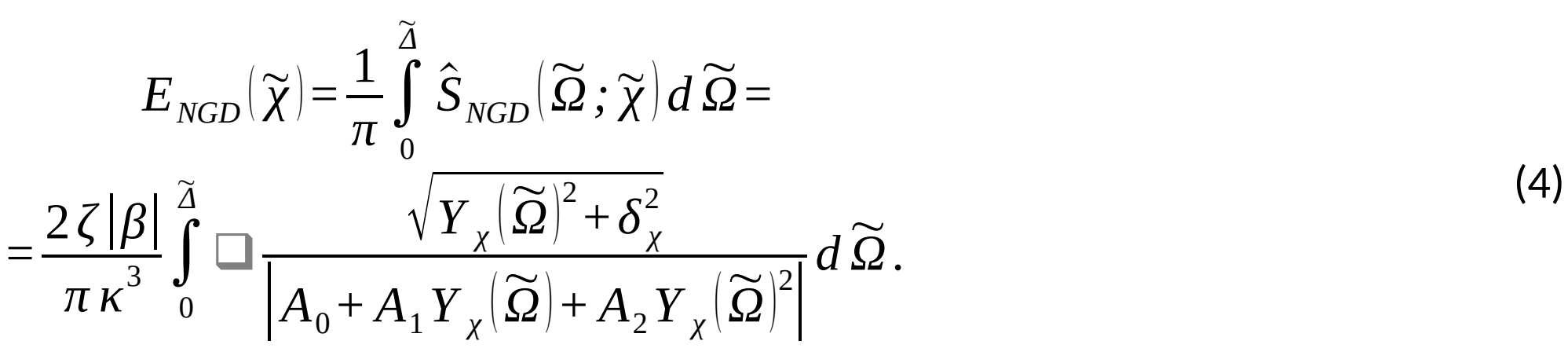

$$E_{NGD}(\tilde{\chi}) = \frac{1}{\pi}\int_0^{\tilde{\Delta}} \hat{S}_{NGD}(\tilde{\Omega};\tilde{\chi})\, d\tilde{\Omega} = \tag{4}$$

$$= \frac{2\zeta|\beta|}{\pi\kappa^3}\int_0^{\tilde{\Delta}} \frac{\sqrt{Y_\chi(\tilde{\Omega})^2+\delta_\chi^2}}{\left|A_0+A_1 Y_\chi(\tilde{\Omega})+A_2 Y_\chi(\tilde{\Omega})^2\right|}\, d\tilde{\Omega}.$$

Let $\theta \equiv Y'_\chi(\tilde{\Omega}_0)$ and set $y := Y_\chi(\tilde{\Omega})$. Using the expansion coefficients $A_{0,1,2}$ defined above, the node–regularized energy takes the compact form

$$E_{NGD}(\chi) = \frac{1}{\pi|\theta|}\int_{y_L}^{y_U} \frac{\sqrt{y^2+\delta_\chi^2}}{\left|A_0+A_1 y+A_2 y^2\right|}\, dy,$$

with $y_L = Y_\chi(0)$ and $y_U = Y_\chi(\tilde{\Delta})$.

4.2.1. Quadratic denominator kept: elementary closed form

Let retain the quadratic term but keep the linearization $y = Y_\chi(\tilde{\Omega}) \approx \theta(\tilde{\Omega} - \tilde{\Omega}_0)$. Set $y = \delta_\chi \sinh\tau$ (so $\sqrt{y^2+\delta_\chi^2} = \delta_\chi \cosh\tau$), and define

$$A = A_2\delta_\chi^2,\, B = A_1\delta_\chi,\, G = A_0,\, D = 4AG - B^2.$$

Then

$$E_{NGD}(\tilde{\chi}) \simeq \frac{2\zeta|\beta|}{\pi\kappa^3|\theta|}\int_{\tau_L}^{\tau_U} \frac{\delta_\chi^2\cosh^2\tau}{A\sinh^2\tau + B\sinh\tau + G}\, d\tau, \quad \left(\tau_{L,U} = arcsinh(y_{L,U}/\delta_\chi)\right). \tag{5}$$

This integral evaluates in elementary functions. The final form depends only on the sign of the discriminant. Let define

$$W(y) = \sqrt{y^2+\delta_\chi^2},$$

then

$$E_{NGD}(\tilde{\chi}) = \frac{2\zeta|\beta|}{\pi\kappa^3|\theta|}\begin{cases}\left[\Phi_{\text{arctan}}(y)\right]_{y_L}^{y_U}, & D>0,\\ \left[\Phi_{\text{artanh}}(y)\right]_{y_L}^{y_U}, & D<0,\end{cases} \tag{6}$$

with

$$\Phi_{\text{arctan}}(y) = \frac{1}{2A}\ln\left(Ay^2+By+G\right) + \frac{B}{2A\sqrt{D}}\, asinh\left(\frac{y}{\delta_\chi}\right) -$$

$$-\frac{1}{\sqrt{D}}\arctan\left(\frac{2AW(y)+By}{\sqrt{D}}\right), (D>0),$$

$$\Phi_{\text{artanh}}(y)=\frac{1}{2A}\ln\left(Ay^2+By+G\right)+\frac{B}{2A\sqrt{-D}}asinh\left(\frac{y}{\delta_\chi}\right)-$$

$$-\frac{1}{\sqrt{-D}}\text{artanh}\left(\frac{2AW(y)+By}{\sqrt{-D}}\right),(D<0).$$

If $D=0$, the equations follow by a continuous limit. That reduces to the arctan–only result above when $A_2\to 0$.

4.3. Expansion of the cutoff branches in powers of the perturbation parameter

In the vicinity of $\tilde{\chi}=0$, the algebraic cut-off branches of the related research article [1] may be expanded in powers of $\tilde{\chi}$ as

$$\tilde{\Delta}\mp^2(\tilde{\chi})=\sum_{i=0}^{\infty}\tilde{F}_i^{\mp}\tilde{\chi}^i, \tag{7}$$

with the first two terms explicitly derived as:

$$\tilde{F}_0^{\mp}=\frac{3}{8}\left[(2-C)\mp\sqrt{(2-C)^2-16\Sigma}\right],$$
$$\tilde{F}_1^{\mp}=\frac{3(2-C)}{8}\pm\frac{3C\Sigma}{8\sqrt{(2-C)^2-16\Sigma}}-\frac{2C-C^2}{4}\tilde{F}_0^{\mp}. \tag{8, 9}$$

The zero-order term corresponds to a strongly chirped DS of the reduced cubic–quintic CGLE in NGD and to the limiting branch points shown in the branch diagram of the related research article [1]. The general recurrence formula is:

$$\tilde{F}_i^{\mp}=\frac{a_i}{8}\mp\sum_{k=0}^{i-1}\binom{1/2}{k}\frac{(-1)^k(16C\Sigma)^k}{\left[(C-2)^2-16\Sigma\right]^{k+1/2}}b_{i-k}-\sum_{j=1}^{min(i,3)}d_j\tilde{F}_{i-j}^{\mp},i\geq 1, \tag{10}$$

with the defined coefficients:

$$\begin{aligned} a_i &=\left[3(2-C),C^2+2C-8+16\Sigma,-16C\Sigma\right], & a_i &=0\text{ for } i\geq 4\\ b_i &=\left[3,\mp(C+4)\right], & b_i &=0\text{ for } i\geq 3\\ d_i &=\left[1,-2(2-C),(2-C)^2\right], & d_i &=0\text{ for } i\geq 4.\end{aligned} \tag{11-13}$$

## 4.4. Saturable gain cavity maps and master-diagram coordinates

In the master-diagram construction (1), $\Sigma$ is treated as an independent control parameter defining a stationary DS existence for given $(C,\tilde{\chi},\Sigma)$.

In a cavity map with saturable gain, however, $\sigma$ is typically *not* an independent parameter. Instead, $\sigma$ is updated self-consistently from the evolving pulse energy each round-trip, for example, by a stiffness law of the form

$$\sigma_{n+1}=(1-r)\sigma_n+r\delta\left(\frac{E_n}{E_{cw}}-1\right), \tag{14}$$

or by a saturable-gain law $g(E)=g_0/(1+E/E_{sat})$ combined with the fixed unsaturated losses. In either case, the steady-state soliton selects a fixed operating point $(E,\sigma)$. Here, $n$ is the round-trip (map-iteration) index, $E_n$ is the pulse energy at round trip $n$ (e.g., $E_n=\int|a_n(t)|^2 dt$ in the same units as $E_{cw}$ or $E_{sat}$). Here, $E_{cw}$ and $E_{sat}$ are the continuous-wave or gain-saturation energy, respectively; $\sigma_n$ is the (saturated) net-loss at round trip $n$ (with $\sigma>0$ meaning a vacuum-stable net loss and $\sigma<0$ meaning a vacuum-unstable net gain). $r\in(0,1]$ is the relaxation/update factor controlling how fast $\sigma_n$ follows the instantaneous value ($r=1$ for instantaneous, smaller $r$ for slower); $\delta$ is the stiffness of the energy-to-loss feedback in $\sigma_n(E)$, and $g_0$ is the small-signal (unsaturated) gain parameter in an explicit saturable-gain model (pump-controlled).

Thus, $\Sigma$ is an output of the steady state rather than a scanned input. As a result, one may observe that $\sigma$ (and therefore $\Sigma$) depends only weakly on the stiffness $\delta$, on the pump proxy, or even moderately on $C$. These parameters primarily change how the map converges, while the stationary pulse balance selects the net-loss level required for equilibrium.

Eq. (2) shows that $\Sigma$ depends on the ratio $\zeta/\kappa$ as well as on $\sigma$. In the gain-saturable simulations, the same coefficients $\kappa,\zeta$ also enter the energy normalization (e.g., through the definition of the saturation energy $E_{sat}$ or its counterpart like $E_{cw}$). As a consequence, the varying $\zeta$, while keeping the cavity-map structure unchanged, changes the mapping between "pump" parameters and the dimensionless operating point. The simulation then readjusts the steady-state energy so that the saturation rule (14) produces a new $\sigma$. It is therefore common to observe partial compensation: $\sigma$ shifts so that the combination $\zeta\sigma/\kappa$ (i.e., $\Sigma$) remains approximately fixed. Thus, the cavity map selects (approximately) one effective isogain slice in a master diagram.

In the master-diagram studies of chirped DSs, the stability/existence regions are often visualized as being "filled" by isogain curves (1): each curve corresponds to a different saturated net-loss level (equivalently, a different pump/unsaturated-loss operating point), and therefore to a different value of $\Sigma$. In a single fixed cavity-map run, only one (or a very narrow set of) operating points is realized, so the numerically observed solutions naturally cluster near some $\Sigma$. To reproduce the filled master-diagram picture numerically, one must introduce an *independent* parameter that shifts the gain–loss balance without being slaved to only the pulse energy. For instance, one may use:

$$\sigma_{n+1}=(1-r)\sigma_n+r\left[\sigma_0+\delta\left(\frac{E_n}{E_{cw}}-1\right)\right], \tag{15}$$

and scan a $\sigma_0$ (an offset or baseline net-loss) at fixed $(\kappa,\zeta,\alpha,\beta,\gamma,\chi)$. This produces a family of steady states with different $\sigma$ and hence different $\Sigma$. Or one may use an explicit saturable-gain model,

$$\sigma(E)=L-\frac{g_0}{1+E/E_{sat}}, \tag{16}$$

and scan the small-signal gain $g_0$ (pump) or/and the unsaturated loss $L$. This allows generating a family of operating points with different $\Sigma$ and populating the master-diagram region. Without such an additional independent knob, a gain-clamped map is expected to yield an approximately single-$\Sigma$ operating line even if $E$ varies strongly.

$\sigma_0$ in Eq. (15) represents the energy-independent part of the net loss relative to threshold (unsaturated intracavity loss minus small-signal gain), while the second term describes the energy-dependent clamping. Experimentally, $\sigma_0$ can be tuned by changing the output coupling, inserting/removing linear attenuation, adjusting intracavity apertures (hard/soft), or varying other baseline loss channels; in SESAM- or Kerr-lens-assisted systems it also effectively captures slow changes of intracavity transmission that are not slaved to the pulse energy on a single round trip. Scanning $\sigma_0$ (or, equivalently, the small-signal gain $g_0$ at fixed unsaturated loss $L$) therefore provides a controlled route to sweep the operating $\Sigma$ across the master diagram, which is essential in the anomalous-dispersion case where the strongly chirped branch exists only inside a finite adiabatic-existence window in $(C,\tilde{\chi},\Sigma)$.

### 4.5. Uniform Airy approximation at the NGD spectral edge

Let us write the (dimensionless) spectral amplitude as a rapidly oscillatory Fourier–type integral

$$\hat{a}(\tilde{\Omega})=\int_{-\infty}^{\infty} A(t)\,e^{\frac{i}{\varepsilon}\Psi(t;\tilde{\Omega})}dt,\ \Psi(t;\tilde{\Omega})=\Phi(t)-\tilde{\Omega}t,\ \Phi'(t)=\tilde{\Omega}_{inst}(t). \tag{17}$$

Here $\varepsilon \ll 1$ is the adiabatic (large-chirp) parameter. With the normalizations (2), it is inversely proportional to the chirp magnitude, so SPA/CFU expansions are controlled by $\varepsilon \to 0$.

Inside the band, the SPA condition $\partial_t \Psi=0$ has two real roots $t_{\pm}(\tilde{\Omega})$, which coalesce at the cut-off $|\tilde{\Omega}|=\tilde{\Delta}$. Thus, the standard two-saddle SPA loses uniformity. Near the turning point $(t_*,\tilde{\Omega}=\tilde{\Delta})$, a Chester–Friedman–Ursell (CFU) change of variables reduces the phase to the Airy canonical cubic (see, e.g., Wong [3]):

$$\begin{gathered}\Psi(t;\tilde{\Omega})=\Psi_*+\frac{\varsigma}{3}s^3-\varsigma^{1/3}\upsilon\, s+O(\varepsilon^{2/3}),\\ \varsigma\equiv\frac{1}{2}\left|\tilde{\Omega}''_{inst}(t_*)\right|,\ \upsilon=\varepsilon^{-2/3}\vartheta(\tilde{\Delta}-\tilde{\Omega}),\ \vartheta=\varsigma^{-2/3}.\end{gathered} \tag{18}$$

Freezing the factor $G(s;\varepsilon)$ at the turning point gives the leading uniform approximation

$$\hat{a}_{unif}(\tilde{\Omega})=e^{i(\Psi_*/\varepsilon+\pi/6)}\left(\frac{2\varepsilon}{\varsigma}\right)^{1/3}\{a_0 Ai(\upsilon)+\varepsilon^{1/3}a_1 Ai'(\upsilon)\}+O(\varepsilon^{4/3}), \tag{19}$$

with $a_0=G_*$ and a computable $a_1$ (a “transport” correction). Keeping only the $Ai$ term already yields a uniform error $O(\varepsilon^{2/3})$ but restores finiteness at the edge.

As $\upsilon \to 0$ (the cut-off), $Ai(0)=3^{-2/3}/\Gamma(2/3)$ gives a finite nonzero edge amplitude. For $\upsilon \to +\infty$ (outside the band), one obtains the exponentially small tail $Ai(\upsilon)\sim\frac{1}{2\sqrt{\pi}}\upsilon^{-1/4}e^{-\frac{2}{3}\upsilon^{3/2}}$. Inside the band ($\upsilon \to -\infty$), the uniform expression reproduces the interference of the two interior stationary points.

With our normalization $\hat{S}(\tilde{\Omega})=P\left|\hat{a}(\tilde{\Omega})\right|^2$ ($P$ as in the main text), a robust one-line uniformization that exactly matches the analytic edge value from the SPA formula is

$$\hat{S}_{unif}(\tilde{\Omega}) = \hat{S}_{edge}\, g\left[\frac{Ai(\upsilon(\tilde{\Omega}))}{Ai(0)}\, g\right]^2, \upsilon(\tilde{\Omega}) = \varepsilon^{-2/3}\vartheta(\tilde{\Delta} - \tilde{\Omega}), \tag{20}$$

with the symmetric form obtained by replacing $(\tilde{\Delta} - \tilde{\Omega})$ by $(\tilde{\Delta} - |\tilde{\Omega}|)$. Here $\hat{S}_{edge}$ is the finite cut-off value from the closed NGD expression. This replacement removes the $0/0$ singularity at $|\tilde{\Omega}| = \tilde{\Delta}$, smooths the cut-off, and agrees with the interior SPA away from the edge.

## 4.6. Coherent multi-horn spectra in AGD

The related research article writes the SPA spectral field as a coherent two-saddle sum with saddle amplitudes $A_\pm$ and phases $\Phi_\pm$. The envelope spectra used there are obtained after averaging the $\pm$ interference term in the power spectrum. In the AGD regime $(\beta < 0, C < 0, \tilde{\chi} > 0)$, this fringe-averaged envelope is denoted below by $\hat{S}_{AGD}^{(env)}(\tilde{\Omega})$.

In fully coherent numerical simulations, or in experiments without sufficient ensemble/time averaging, the $\pm$ interference is not removed. Retaining it gives the coherent (non-averaged) spectral power in the standard two-saddle form

$$\hat{S}_{AGD}^{(coh)}(\tilde{\Omega}) = \hat{S}_{AGD}^{(env)}(\tilde{\Omega})\left[1 + V(\tilde{\Omega})\cos\Delta\Phi(\tilde{\Omega})\right],$$
$$V(\tilde{\Omega}) = \frac{2\left|A_+(\tilde{\Omega})\,A_-(\tilde{\Omega})\right|}{\left|A_+(\tilde{\Omega})\right|^2 + \left|A_-(\tilde{\Omega})\right|^2} \in [0,1], \tag{21}$$

where $\hat{S}_{AGD}^{(env)}(\tilde{\Omega})$ is the fringe-averaged AGD SPA envelope, $A_\pm$ are the two saddle amplitudes, and $\Delta\Phi(\tilde{\Omega}) \equiv \Phi_+(\tilde{\Omega}) - \Phi_-(\tilde{\Omega})$ is the relative saddle phase computed from the same AGD chirp law.

For a symmetric single-hump pulse, one typically has $|A_+| \simeq |A_-|$ in the central band, hence $V(\tilde{\Omega}) \simeq 1$ there. While in the far wings one saddle dominates and $V(\tilde{\Omega}) \to 0$, so that $\hat{S}_{AGD}^{(coh)} \to \hat{S}_{AGD}^{(env)}$ asymptotically. The interference factor in Eq. (21) can therefore (i) preserve a two-horn profile when $\cos\Delta\Phi$ is destructive near $\tilde{\Omega} \simeq 0$, or (ii) fill the smooth central depression of the envelope and produce a stable three-horn pattern when the interference becomes sufficiently constructive in the central band. Because both $V(\tilde{\Omega})$ and $\Delta\Phi(\tilde{\Omega})$ vary smoothly with the parameters, the transition between two- and three-horn spectra is continuous.

The horn positions are governed by the constructive-interference condition $\Delta\Phi(\tilde{\Omega}) = 2\pi m$ (local maxima for $V$ are not too small), while the local minima satisfy $\Delta\Phi(\tilde{\Omega}) = (2m+1)\pi$. This coherent-fringe mechanism is independent of the NGD node regularization discussed above.

## 4.7. Dissipative-soliton energy in AGD

For the AGD branch we use the positive scale

$$D_a \equiv -\tilde{\Delta}_-^2 > 0, b \equiv 1 + 4\tilde{\chi}D_a > 1.$$

The radical-free parametrization is

$$Y=\sqrt{b}\cosh\tau\,,\tilde{\Omega}^2=\frac{b}{4\tilde{\chi}}\sinh^2\tau\,,d\tilde{\Omega}=\frac{\sqrt{b}}{2\sqrt{\tilde{\chi}}}\cosh\tau\,d\tau\,.$$

The function $G(\tilde{\Omega})$ entering the AGD stationary-phase spectrum can be written as a quadratic polynomial in $Y$,

$$G(\tilde{\Omega})=p_2Y^2+p_1Y+p_0,$$

where

$$p_2=\frac{1+C\tilde{\chi}}{2},p_1=\tilde{\chi}-1,p_0=\frac{1}{2}-\tilde{\chi}-\frac{C\tilde{\chi}}{2}b+2\tilde{\chi}^2\Sigma.$$

Therefore,

$$\tilde{L}(\tau)\equiv G(\tilde{\Omega}(\tau))=\Lambda_2\cosh^2\tau+\Lambda_1\cosh\tau+\Lambda_0,$$

with

$$\Lambda_2=p_2b\,,\Lambda_1=p_1\sqrt{b}\,,\Lambda_0=p_0.$$

We also define

$$D_\Lambda\equiv4\Lambda_2\Lambda_0-\Lambda_1^2.$$

We start from the AGD energy representation used in the related research article [1]:

$$E_{AGD}=\frac{2\zeta|\beta|}{\pi\kappa^3}[\Psi(\tau)]_{\tau=0}^{\infty},$$

$$\Psi(\tau)=\Phi_{\log}(\tau)+\begin{cases}\frac{A_1}{\sqrt{D_\Lambda}}\arctan\left(\frac{2\Lambda_2\sinh\tau+\Lambda_1}{\sqrt{D_\Lambda}}\right), & D_\Lambda>0,\\ \frac{A_1}{\sqrt{-D_\Lambda}}arctanh\left(\frac{2\Lambda_2\sinh\tau+\Lambda_1}{\sqrt{-D_\Lambda}}\right), & D_\Lambda<0,\\ A_1\frac{2\Lambda_2\sinh\tau+\Lambda_1}{\Lambda_2\sinh^2\tau+\Lambda_1\sinh\tau+\Lambda_0}, & D_\Lambda=0,\end{cases} \quad (22)$$

where the logarithmic part is

$$\Phi_{\log}(\tau)=\frac{A_0}{2\Lambda_2}\ln\tilde{L}(\tau)-\frac{B_0}{2b}\ln(b\cosh^2\tau-1)+\frac{B_1}{\sqrt{b-1}}arctanh\left(\frac{\sqrt{b}\sinh\tau}{\sqrt{b-1}}\right).$$

The constants $A_0,B_0,A_1,B_1,A_2$ are fixed by the partial-fraction identity

$$\frac{R(\cosh\tau)}{(b\cosh^2\tau-1)\tilde{L}(\tau)^2}=\frac{d}{d\tau}\left[\frac{A_0}{2\Lambda_2}\ln\tilde{L}(\tau)-\frac{B_0}{2b}\ln(b\cosh^2\tau-1)+I_\Lambda(\tau)+\frac{B_1}{\sqrt{b-1}}arctanh\left(\frac{\sqrt{b}\sinh\tau}{\sqrt{b-1}}\right)+A_2\tau\right],$$

where

$$R(\cosh\tau)=\frac{\sqrt{\tilde{\chi}}}{2}\left(2b\cosh^2\tau-b-\sqrt{b}\cosh\tau\right).$$

The term $I_\Lambda(\tau)$ depends on the sign of $D_\Lambda$:

$$I_\Lambda(\tau)=\begin{cases}\dfrac{A_1}{\sqrt{D_\Lambda}}\arctan\left(\dfrac{2\Lambda_2\cosh\tau+\Lambda_1}{\sqrt{D_\Lambda}}\right), & D_\Lambda>0,\\ \dfrac{A_1}{\sqrt{-D_\Lambda}}arctanh\left(\dfrac{2\Lambda_2\cosh\tau+\Lambda_1}{\sqrt{-D_\Lambda}}\right), & D_\Lambda<0,\\ \dfrac{A_1}{2\Lambda_2\cosh\tau+\Lambda_1}, & D_\Lambda=0.\end{cases}$$

Equivalently, the constants $A_0, B_0, A_1, B_1, A_2$ are obtained by differentiating the expression in brackets and matching the coefficients of equal powers of $\cosh\tau$. This gives purely algebraic functions of $(\tilde{\chi}, C, \Sigma, D_a)$.

Let $W(\tau)\equiv\sqrt{b}\sinh\tau$. Then, the primitive can be written as

$$\Psi(\tau)=\frac{A_0}{2\Lambda_2}\ln\tilde{L}(\tau)-\frac{B_0}{2b}\ln\left(b\cosh^2\tau-1\right)+ \\ +I_\Lambda(\tau)+\frac{B_1}{\sqrt{b-1}}arctanh\left(\frac{\sqrt{b}\sinh\tau}{\sqrt{b-1}}\right)+A_2\tau. \quad (23)$$

with constants $A_0, B_0, A_1, B_1, A_2$ given explicitly by the (lengthy) algebraic combinations of $\tilde{\chi}, C, \Sigma, D_a$ that arise in the partial-fraction decomposition of $R/\left[\left(b\cosh^2\tau-1\right)\tilde{L}^2\right]$. Since $\cosh\tau\to 1$ as $\tau\to 0$ and $\cosh\tau\sim\frac{1}{2}e^\tau$ as $\tau\to\infty$, all terms have finite limits and the energy is

$$E_{AGD}=\frac{2\zeta|\beta|}{\pi\kappa^3}\{\Psi(\infty)-\Psi(0)\}. \quad (24)$$

## 4.8. Autocorrelation function for a strongly chirped AGD dissipative soliton

### 4.8.1. Definitions and windowed spectrum

In the AGD regime $\left(\beta<0, C<0, \tilde{\chi}>0, \tilde{\Delta}_-^2<0\right)$, the SPA yields the envelope spectrum $\hat{S}_{AGD}(\tilde{\Omega})$ used in the related research article [1]. This spectrum has no strict cut-off, decays as $\hat{S}_{AGD}(\tilde{\Omega})\sim|\tilde{\Omega}|^{-3}$, and therefore has a finite energy.

In practice, finite spectral dissipation, measurement bandwidth, or the finite plotting window used for the AGD spectra in the related research article [1] produces a windowed spectrum

$$\hat{S}_{AGD}^{(cap)}(\tilde{\Omega})=\hat{S}_{AGD}(\tilde{\Omega})H\left(\tilde{\Omega}_{cap}-|\tilde{\Omega}|\right), \quad (25)$$

where $H$ is the Heaviside function and $\tilde{\Omega}_{cap}$ is the effective observation bandwidth.

The (first-order) field autocorrelation is defined by the cosine transform

$$R^{(1)}_{cap}(\tilde{\tau})=\frac{1}{\pi}\int_0^{\infty}\hat{S}^{(cap)}_{AGD}(\tilde{\Omega})\cos(\tilde{\Omega}\tilde{\tau})\,d\tilde{\Omega},\; g^{(1)}_{cap}(\tilde{\tau})=\frac{R^{(1)}_{cap}(\tilde{\tau})}{R^{(1)}_{cap}(0)}. \tag{26}$$

(If the field fluctuations are close to Gaussian, the intensity autocorrelation follows from a Siegert-type relation: $g^{(2)}(\tilde{\tau})=1+|g^{(1)}(\tilde{\tau})|^2$.)

### 4.8.2. Convolution form and the short correlation scale

Let $F^{-1}$ denote the inverse Fourier transform in $\tilde{\Omega}$. Since multiplication in frequency corresponds to convolution in time, Eq. (25) implies

$$R^{(1)}_{cap}(\tilde{\tau})=R^{(1)}_{\infty}(\tilde{\tau})*K_{cap}(\tilde{\tau}),\; K_{cap}(\tilde{\tau})=\frac{\tilde{\Omega}_{cap}}{\pi}sinc(\tilde{\Omega}_{cap}\tilde{\tau}), \tag{27}$$

where $R^{(1)}_{\infty}$ is the correlation corresponding to the unwindowed spectrum $\hat{S}_{AGD}$ and $sinc(x)\equiv\sin(x)/x$. Thus, the spectral windowing introduces a *short* correlation scale

$$l\sim\frac{\pi}{\tilde{\Omega}_{cap}}, \tag{28}$$

which is the characteristic width of $K_{cap}$. This is the direct AGD analogue of the “$sinc$-graining” mechanism and the microscopic time scale identified for strongly chirped DSs in Appendix B of Ref. [2].

### 4.8.3. Long correlation scale from the spectral core

The *long* correlation scale is controlled by the spectral core of $\hat{S}_{AGD}$. A convenient operational definition is to introduce a core bandwidth $\tilde{\Omega}_{core}$ by an energy fraction

$$\int_0^{\tilde{\Omega}_{core}}\hat{S}_{AGD}(\tilde{\Omega})\,d\tilde{\Omega}=(1-\varepsilon)\int_0^{\infty}\hat{S}_{AGD}(\tilde{\Omega})\,d\tilde{\Omega},\; 0<\varepsilon\ll 1, \tag{29}$$

and set

$$\varrho\sim\tilde{\Omega}^{-1}_{core}. \tag{30}$$

Then $|g^{(1)}_{cap}(\tilde{\tau})|$ typically exhibits a broad pedestal of width $\sim\varrho$, while the $sinc$-kernel in Eq. (27) produces a much narrower structure of width $\sim l$, so that a two-scale separation $l\ll\varrho$ emerges.

For the analytic estimations, one may approximate the core of the AGD spectrum by a Lorentzian $\hat{S}_{core}(\tilde{\Omega})\approx A(1+\tilde{\Omega}^2/\Gamma^2)^{-1}$, which gives $R_{core}(\tilde{\tau})\propto e^{-\Gamma|\tilde{\tau}|}$ and hence $\varrho\sim 1/\Gamma$. Substitution into Eq. (27) yields the explicit two-scale convolution

$$R^{(1)}_{cap}(\tilde{\tau})\approx\int_{-\infty}^{\infty}e^{-|t|/\varrho}\frac{\tilde{\Omega}_{cap}}{\pi}sinc(\tilde{\Omega}_{cap}(\tilde{\tau}-t))\,dt, \tag{31}$$

which is structurally identical to the NGD strong-chirp autocorrelation formula of Ref. [2].

### 4.8.4. Contribution of the algebraic AGD tail

The AGD envelope has an algebraic wing $\hat{S}_{AGD}(\tilde{\Omega}) \sim K/|\tilde{\Omega}|^3$ with

$$K \propto \frac{1}{(1+C\tilde{\chi})^2\sqrt{\tilde{\chi}}}, \tag{32}$$

as follows from the AGD asymptotic form used in the related research article [1]. Let $\tilde{\Omega}_0$ denote a matching frequency beyond which the $|\tilde{\Omega}|^{-3}$ asymptotic is accurate.

The tail contribution to $R^{(1)}_{cap}$ is then

$$R_{wing}(\tilde{\tau}) \approx \frac{K}{\pi}\int_{\tilde{\Omega}_0}^{\tilde{\Omega}_{cap}} \frac{\cos(\tilde{\Omega}\tilde{\tau})}{\tilde{\Omega}^3} d\tilde{\Omega}. \tag{33}$$

Expanding $\cos(\tilde{\Omega}\tilde{\tau})$ for $|\tilde{\tau}| \ll \tilde{\Omega}^{-1}_{cap}$ gives

$$R_{wing}(\tilde{\tau}) = R_{wing}(0) - \frac{K}{2\pi}\tilde{\tau}^2 \ln\left(\frac{\tilde{\Omega}_{cap}}{\tilde{\Omega}_0}\right) + O(\tilde{\tau}^4), \tag{34}$$

showing that the curvature of the correlation peak depends logarithmically on the spectral cut-off. Near the AGD–DSR line $1+C\tilde{\chi} \simeq 0$, the prefactor $K$ increases, enhancing the relative weight of high-frequency components and, thus, amplifying the short-scale structure (i.e., increasing the separation between $l$ and $\varrho$).

### 4.8.5. Note on thermodynamic interpretation

A heuristic "microstate" interpretation of the scale separation $l \ll \varrho$ and its connection to multi-pulse tendencies is discussed in the related research article [1] and in Ref. [2]. It is omitted here to keep the present data article focused on datasets and reproducibility.

### 4.9. MATLAB stability-analysis data for AGD dissipative solitons

The repository contains the MATLAB scripts and processed data used to generate the AGD finite-time survival maps reported in the related research article [1] and reproduced here in Figure 1. These calculations test the dynamical accessibility of a prepared analytical strongly chirped AGD branch under quantum-noise-level perturbations. Each parameter point is initialized from the reconstructed analytical pulse and propagated with an independent band-limited Wigner-vacuum perturbation.

The numerical procedure is as follows. First, the analytical AGD field a0(t) is reconstructed from the stationary-phase solution. Second, a small complex perturbation δa is initialized by independent complex Gaussian spectral amplitudes inside the effective spectral window. Third, the linearized Bogoliubov-type equation is integrated by a split-step method: spectral filtering and GDD are applied in the Fourier domain, while the local 2×2 coupling between δa and δa* is evaluated in the time domain. The propagation is performed in the natural Fourier-conjugate variables (t, Ω); normalized coordinates are used for presentation and tabulation.

For the stability maps, each grid point uses Nshot = 32 independent noise realizations over the propagation interval $0 < z < 200$. A realization is terminated if the normalized perturbation distance reaches εmax = 0.25. It is counted as surviving only if it remains single-pulse over the interval, does not reach the perturbation cutoff, and retains high spectral-shape correlation with the analytical reference. The map value is Psurv = Nsurv/Nshot, and Psurv > 0.5 is used as the practical finite-time survival criterion.

The maps use $\tilde{\chi}$= 1.5, κ/γ = 0.10, ζ/γ = 0.39, and α = Cβκ/γ with C = -1/$\tilde{\chi}$+ δC. Figure 1(a) scans the (Ẽ, δC) plane at fixed Σ = 0.14, whereas Figure 1(b) scans the (Σ, δC) plane at fixed Ẽ = 8. The representative-profile histograms in Figure 2 use 64 independent noise realizations at a selected point inside the finite-time-survival region; this larger ensemble is distinct from the 32-shot ensemble used at each map pixel.

The temporal-grid size, Fourier-window extent, propagation step, and effective noise bandwidth are script-level discretization parameters stored with the deposited MATLAB source. Reproduction of the published maps should use those archived values unchanged; varying them constitutes the grid/window/step convergence test described in Section 5.

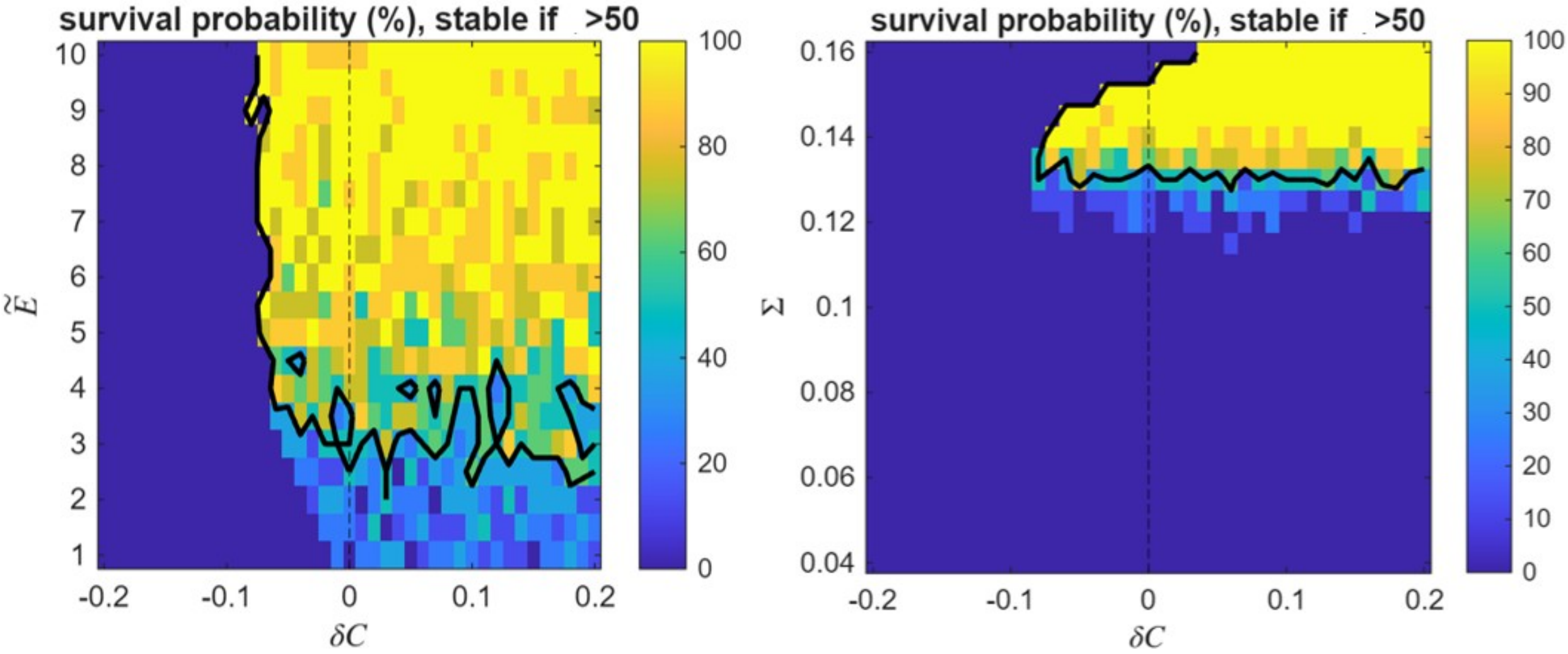

Figure 1. Stability regions of the analytical strongly chirped AGD-DS under quantum-noise perturbations. The stability criterion is $P_{_surv} > 50\%$, where $P_{_surv}$ is the fraction of noisy realizations that remain single-pulse, do not reach the perturbation cutoff ||δa||/||a_0||=0.25, and retain high spectral correlation with the analytical reference over the propagation interval. Yellow denotes the stable/long-lived region, and blue denotes unstable points. The vertical dashed line marks the AGD--DSR condition $\delta C = 0$, where $C = -1/\tilde{\chi} + \delta C$. Parameters common to both panels are $\tilde{\chi} = 1.5$, $\kappa/\gamma = 0.1$, $\zeta/\gamma = 0.39$, and $\alpha = C\beta\kappa/\gamma$ (see Eqs. (1,2)). (a): Stability map in the $\tilde{E}, \delta C$ plane at fixed $\Sigma = 0.14$. (b) Stability map in the $\Sigma, \delta C$ plane at fixed $\tilde{E} = 8$.

The exported stability tables record the scanned control parameters together with Psurv and the retained diagnostics for each parameter point. The 50% criterion is a finite-time robustness classification under the specified perturbation model; it should not be interpreted as an asymptotic orbital-stability boundary.

The representative-profile files reproduce the comparison between the analytical pulse and a noisy final state shown in Figure 2. They contain the analytical table-top temporal envelope, the final noisy envelope, the corresponding two-horn AGD spectral core, the distribution of spectral correlations over 64 independent realizations, and the associated dimensionless-energy distribution.

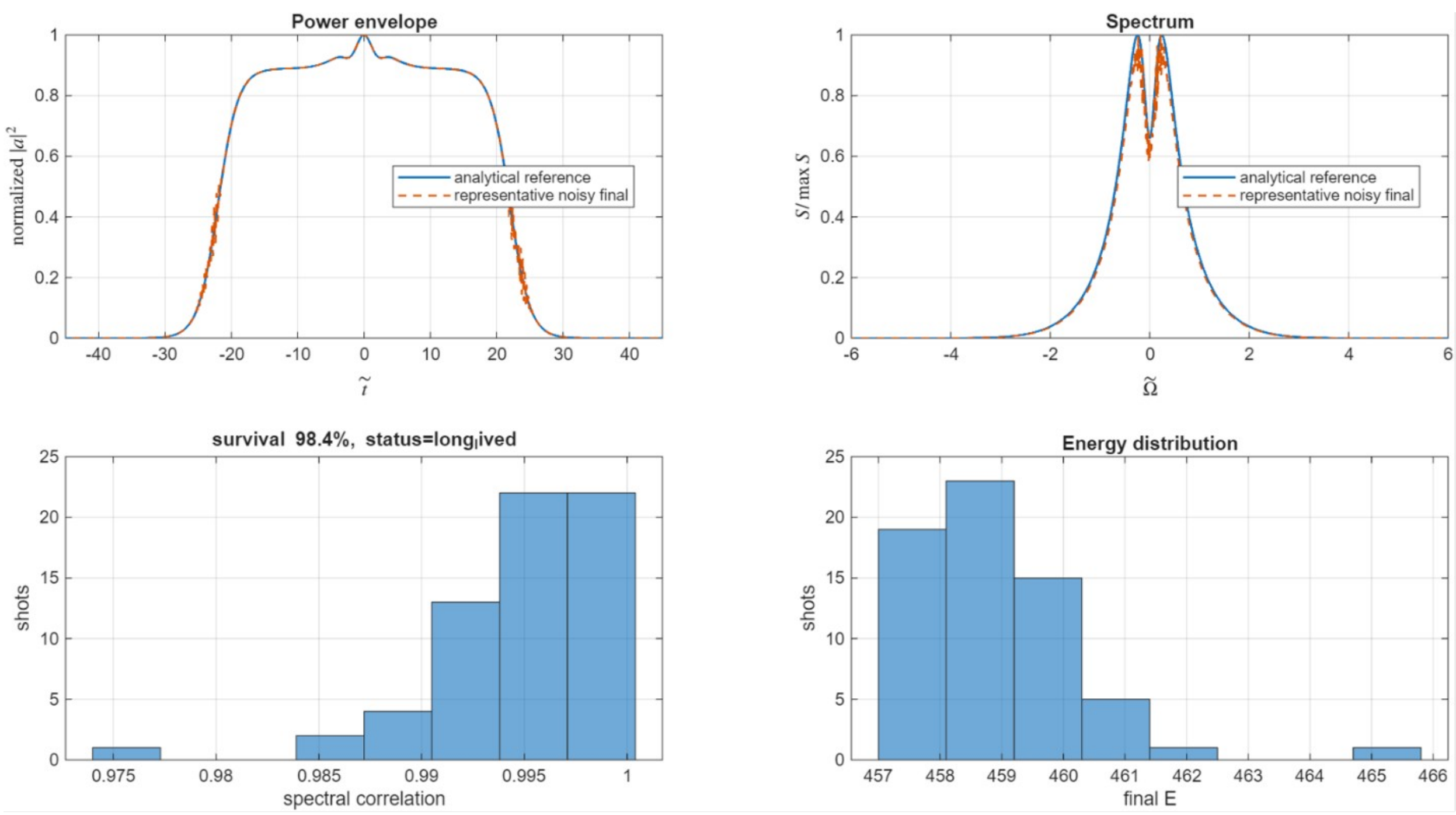

Figure 2. DS power envelope (upper-left) and spectral profile (upper-right) for analytical solution (solid blue) and noisy perturbed one (dashed red). The bottom-left histogram demonstrates the spectral correlation of the shots with high $P_{surv}$ from 64 independent noise realizations. The bottom-right histogram demonstrates the corresponding dimensionless energy distribution. The dimensionless energy equals $\tilde{E}=8, C=-0.67, \delta C=-0.01, \Sigma=0.14, \tilde{\chi}=1.5$. Other parameters correspond to Figure 1.

## 5. Technical Validation

Validation was performed at three levels. Analytical consistency was checked by enforcing the branch reality and admissibility conditions and by comparing limiting cases, including χ̃-> 0, with the reduced-CGLE expressions. Numerical spectra and correlation functions were checked by repeating selected evaluations on refined grids and by cross-comparing FFT/cosine-transform results with direct quadrature. Convergence is judged operationally by invariance of the plotted profile and of the relevant integrated quantities at the precision reported in the exported tables; the script-level grid, window, and step parameters are retained with the deposited source so that these checks can be repeated.

The AGD map is a finite-time stochastic classification rather than an asymptotic stability calculation. With Nshot = 32 at each map point, the binomial standard error at Psurv = 0.5 is approximately 0.088; therefore the 50% contour should be interpreted at the resolution of the ensemble and parameter grid. The representative-profile dataset uses 64 independent realizations to characterize the distribution of spectral correlation and energy at a selected robust point.

As a final reproducibility check, the open .csv/.txt exports contain the same numerical columns as the corresponding OriginLab datasets. Re-plotting these tables independently provides a direct check that the archived numerical data, rather than software-specific project state, determine the reported curves and maps.

# LIMITATIONS

The datasets and scripts are limited to the parameter ranges, asymptotic assumptions, and numerical grids described in the repository and in the related research article [1]. The adiabatic strongly chirped approximation, stationary-phase treatment, and windowed-transform calculations should be used within the stated existence and admissibility domains of the cubic-quintic CGLE model.

# ETHICS STATEMENT

Not applicable. This data article reports symbolic derivations, numerical scripts, and processed computational datasets only; it does not involve human participants, animal experiments, clinical data, or personal data.

# CRediT AUTHOR STATEMENT

V. L. Kalashnikov: Conceptualization, Methodology, Software, Formal analysis, Data curation, Writing - original draft, Visualization. E. Sorokin: Methodology, Validation, Resources, Writing - review & editing. A. Rudenkov: Software, Validation, Formal analysis, Writing - review & editing. I. T. Sorokina: Supervision, Project administration, Funding acquisition, Resources, Writing - review & editing.

# ACKNOWLEDGEMENTS

This work was supported by Norges Forskningsråd (#303347 (UNLOCK), #326503 (MIR)) , and ATLA Lasers AS.

# DECLARATION OF COMPETING INTERESTS

The authors declare that they have no known competing financial interests or personal relationships that could have appeared to influence the work reported in this paper.

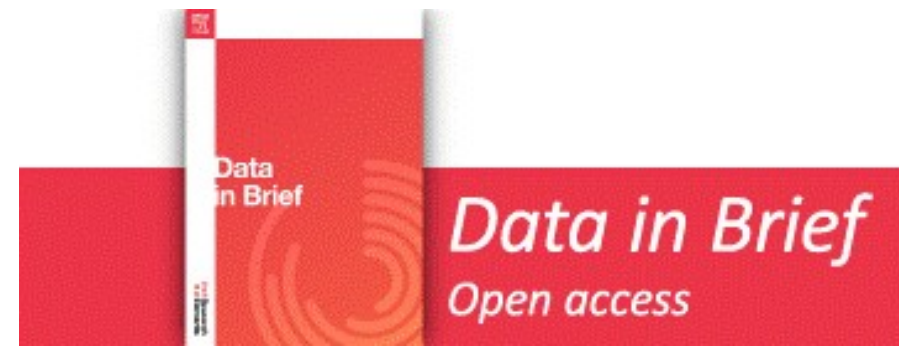
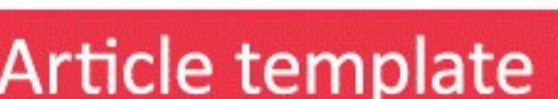